\documentclass[aps,prd,superscriptaddress,twocolumn]{revtex4-1}
\usepackage{amssymb}
\usepackage{amsmath}
\usepackage{wallpaper}
\usepackage{bm}
\usepackage{multirow}
\usepackage[colorlinks,linkcolor=blue, urlcolor=blue, anchorcolor=blue, citecolor=blue]{hyperref}

\begin{document}

\title{Quarkyonic matter and quarkyonic stars in an extended RMF model}

\author{Cheng-Jun Xia}
\email{cjxia@yzu.edu.cn}
\affiliation{Center for Gravitation and Cosmology, College of Physical Science and Technology, Yangzhou University, Yangzhou 225009, China}

\author{Hao-Miao~Jin}
\affiliation{Center for Gravitation and Cosmology, College of Physical Science and Technology, Yangzhou University, Yangzhou 225009, China}
\affiliation{School of Physics and Microelectronics, Zhengzhou University, Zhengzhou 450001, China}

\author{Ting-Ting~Sun}
\affiliation{School of Physics and Microelectronics, Zhengzhou University, Zhengzhou 450001, China}

\date{\today}

\begin{abstract}
By combining RMF models and equivparticle models with density-dependent quark masses, we construct explicitly ``a quark Fermi Sea'' and ``a baryonic Fermi surface'' to model the quarkyonic phase, where baryons with momentums ranging from zero to Fermi momentums are included. The properties of nuclear matter, quark matter, and quarkyonic matter are then investigated in a unified manner, where quarkyonic matter is more stable and energy minimization is still applicable to obtain the microscopic properties of dense matter. Three different covariant density functionals TW99, PKDD, and DD-ME2 are adopted in our work, where TW99 gives satisfactory predictions for the properties of nuclear matter both in neutron stars and heavy-ion collisions and quarkyonic transition is unfavorable. Nevertheless, if PKDD with larger slope of symmetry energy $L$ or DD-ME2 with larger skewness coefficient $J$ are adopted, the corresponding EOSs are too stiff according to both experimental and astrophysical constraints. The situation is improved if quarkyonic transition takes place, where the EOSs become softer and can accommodate various experimental and astrophysical constraints.
\end{abstract}

\maketitle

\section{\label{sec:intro}Introduction}
Due to the asymptotic freedom and confinement of strong interactions at large and small energy scales, the strongly interacting matter at zero temperature is believed to exhibit at least two phases, i.e., low-density hadronic matter (HM) and high-density quark matter (QM). As density increases, HM may undergo a deconfinement phase transition and form QM, while it is not clear what exactly happens in the processes of deconfinement phase transition and many possibilities exist~\cite{Fukushima2005_PRD71-034002, Voskresensky2023_PPNP130-104030}. For example, one type of deconfinement phase transitions from HM to QM are of first-order~\cite{Dexheimer2010_PRC81-045201}, indicating the possible existence of quark-hadron mixed phase inside hybrid stars~\cite{Heiselberg1993_PRL70-1355, Glendenning2001_PR342-393, Voskresensky2002_PLB541-93, Tatsumi2003_NPA718-359, Voskresensky2003_NPA723-291, Endo2005_NPA749-333, Maruyama2007_PRD76-123015, Peng2008_PRC77-065807, Yasutake2014_PRC89-065803, Xia2019_PRD99-103017, Maslov2019_PRC100-025802, Xia2020_PRD102-023031}. The other type of deconfinement phase transitions resemble those at vanishing chemical potentials and large temperatures, where a smooth crossover between HM and quark-gluon plasma takes place~\cite{Borsanyi2014_PLB730-99, Bazavov2014_PRD90-094503}. The hadron-quark crossover at finite densities was modeled with various phenomenological interpolation functions, which predicts stiffer equation of state (EOS) so that a hybrid star could reaches 2 $M_\odot$~\cite{Baym1979_PA96-131, Celik1980_PLB97-128, Schaefer1999_PRL82-3956, Fukushim2004_PLB591-277, Hatsuda2006_PRL97-122001, Maeda2009_PRL103-085301, Masuda2013_ApJ764-12, Masuda2013_PTEP2013-073D01, Zhao2015_PRD92-054012, Kojo2015_PRD91-045003, Masuda2016_EPJA52-65, Whittenbury2016_PRC93-035807, Bai2018_PRD97-023018, Baym2019_ApJ885-42}. The implications of such a hadron-quark crossover on binary neutron star mergers and the postmerger gravitational signals were then examined, which could be identified by future kilohertz gravitational wave detectors~\cite{Huang2022_PRL129-181101, Fujimoto2023_PRL130-91404}.

To unveil the microscopic dynamics for the crossover domain, as demonstrated by Fukushima and Kojo~\cite{Fukushima2016_ApJ817-180}, the crossover from HM to QM can be bridged by quarkyonic matter. At supranuclear densities, the many-body interactions between baryons become significant~\cite{Hebeler2010_PRL105-161102}, which is attributed to the increasing number of exchanged quarks~\cite{Fukushima2016_ApJ817-180}. The boundary between baryons eventually becomes blurred and quarks can move freely among baryons at large densities, i.e., forming the quarkyonic phase~\cite{McLerran2009_NPA830-709c}. As proposed by McLerran and Pisarski in the large $N_\mathrm{c}$ limit~\cite{McLerran2007_NPA796-83}, a quarkyonic phase is comprised of ``a quark Fermi Sea'' and ``a baryonic Fermi surface''. Further studies on the phase diagram of strongly interacting matter with an extended Nambu-Jona-Lasinio model suggest that the quarkyonic transition is indeed a crossover at $N_\mathrm{c}=3$~\cite{McLerran2009_NPA824-86}. It was shown that the pressure and sound velocity of quarkyonic matter increase rapidly with density, which fulfills the observational constraints on massive neutron stars~\cite{McLerran2019_PRL122-122701}. The effects of isospin-flavor asymmetry was later considered, predicting a lower proton fraction which could potentially quench fast cooling in massive quarkyonic stars~\cite{Margueron2021_PRC104-055803}. By synthesizing the Walecka model together with the quark-meson model, a complete field model for quarkyonic matter treating baryons, quarks, and mesons on the same footing was developed~\cite{Cao2020_JHEP10-168, Cao2022_PRD105-114020}, where the chiral symmetry breaking and restoration in quarkyonic matter can be described. In the framework of a constituent quark model, the emergence of quarkyonic modes in dense baryonic matter were analyzed as well, where the attractive $ud$ color antitriplet diquark remains intact~\cite{Park2021_PRD104-094024, Park2022_PRD105-114034}.

For baryonic matter below and around the nuclear saturation density $n_{0}$, relativistic mean field (RMF) models are often adopted~\cite{Meng2016_RDFNS}, which give excellent description for finite nuclei~\cite{Reinhard1989_RPP52-439, Ring1996_PPNP37_193-263, Meng2006_PPNP57-470, Paar2007_RPP70-691, Meng2015_JPG42-093101, Typel1999_NPA656-331, Vretenar1998_PRC57-R1060, Lu2011_PRC84-014328, Roca-Maza2011_PRC84-054309, Belvedere2012_NPA883-1, Chen2021_SCPMA64-282011} and nuclear matter~\cite{Glendenning2000, Ban2004_PRC69-045805, Weber2007_PPNP59-94, Long2012_PRC85-025806, Sun2012_PRC86-014305, Wang2014_PRC90-055801, Fedoseew2015_PRC91-034307, Gao2017_ApJ849-19, Sun2018_CPC42-25101}. In such cases, it is nature to extend RMF models to include quark degrees of freedom, which was done by combining RMF models and equivparticle models with density-dependent quark masses~\cite{Xia2018_JPSCP20-011010}. In this work, based on our previous study~\cite{Xia2018_JPSCP20-011010}, we construct explicitly ``a quark Fermi Sea'' and ``a baryonic Fermi surface'' to model the quarkyonic phase. It is worth mentioning that in contrast to previous constructions of the Fermi sphere for quarkyonic matter by simply removing lower momentum components~\cite{McLerran2019_PRL122-122701, Margueron2021_PRC104-055803, Cao2020_JHEP10-168, Cao2022_PRD105-114020}, baryons with momentums ranging from zero to Fermi momentums are considered here. We believe such treatment is more nature since the low-energy excitations should carry vanishing momentums as in analogous to the formation of Cooper pairs~\cite{Kojo2021_PRD104-63036}, which is dominated by zero momentum components.

The interaction between baryons are treated with the RMF approach via exchange of $\sigma$, $\omega,$ and $\rho$ mesons, where the baryon-meson couplings are density dependent adopting the effective interactions TW99~\cite{Typel1999_NPA656-331}, PKDD~\cite{Long2004_PRC69-034319}, and DD-ME2~\cite{Lalazissis2005_PRC71-024312}. The quarks are considered as quasi-free particles with density dependent masses, including confinement and leading-order perturbative interactions~\cite{Xia2014_PRD89-105027}. Finally, the quark-baryon interactions are accounted for with density dependent baryon masses, and energy minimization is still applicable to obtain microscopic properties of quarkyonic matter.
The paper is organized as follows. In Section~\ref{sec:the}, we present the theoretical framework for nuclear matter, quark matter, and quarkyonic matter. The properties of dense matter and the implication for compact star structures are then examined in Section~\ref{sec:res}. We draw our conclusion in Section~\ref{sec:con}.

\section{\label{sec:the}Theoretical framework}
The Lagrangian density of the extended RMF model can be divided into the following three parts as
\begin{equation}
 \mathcal{L} = \mathcal{L}^\mathrm{B}+ \mathcal{L}^\mathrm{Q}+ \mathcal{L}^\mathrm{L},\label{eq:2.1}
\end{equation}
where $\mathcal{L}^\mathrm{B}$, $\mathcal{L}^\mathrm{Q}$, and $\mathcal{L}^\mathrm{L}$ are respectively the Lagrangian densities for nuclear matter, quark matter, and leptonic matter, i.e.,
\begin{eqnarray}
\mathcal{L}^\mathrm{B} &=& \sum_{i=n,p} \bar{\Psi}_i\{i \gamma^{\mu} \partial_\mu -m_i (n^\mathrm{Q}_\mathrm{b})-g_{\sigma i}(n^\mathrm{B}_\mathrm{b}) \sigma \nonumber \\
  && -g_{\omega i}(n^\mathrm{B}_\mathrm{b})\gamma^\mu\omega_{\mu}-g_{\rho i}(n^\mathrm{B}_\mathrm{b})\gamma^\mu\bm{\tau_i}\cdot\bm{\rho}_{\mu}\} \Psi_{i} \nonumber\\
  && -\frac{1}{2}m_{\sigma}^{2} \sigma^2+\frac{1}{2}m_{\omega}^{2} \omega_{\mu}\omega^{\mu}+\frac{1}{2}m_{\rho}^{2}\bm{\rho}_{\mu} \cdot \bm{\rho}^{\mu}, \label{eq:2.2}\\
\mathcal{L}^\mathrm{Q}&=&\sum_{i=u,d} \bar{\Psi}_{i}[i \gamma^{\mu} \partial _{\mu}-m_{i} (n_\mathrm{b})] \Psi_{i},\label{eq:2.3}\\
\mathcal{L}^\mathrm{L}&=&\sum_{i=e,\mu} \bar{\Psi}_{i}[i \gamma^{\mu} \partial _{\mu}-m_{i}]\Psi_{i} .\label{eq:2.4}
\end{eqnarray}%
Here $\Psi_{i}$ represents the Dirac spinor for different fermions $i$ (baryons, quarks and leptons) with masses $m_{i}$, where $m_{n,p}(n^\mathrm{Q}_\mathrm{b})$ and $m_{u,d}(n_\mathrm{b})$ are density dependent with $n^\mathrm{B}_\mathrm{b}$, $n^\mathrm{Q}_\mathrm{b}$, and $n_\mathrm{b}$ being respectively the baryon number densities for nucleons, quarks, and both particles combined, i.e.,
\begin{equation}
  n^\mathrm{B}_\mathrm{b}=n_p + n_n;\  n^\mathrm{Q}_\mathrm{b}=(n_u + n_d)/3;\  n_\mathrm{b}=n^\mathrm{B}_\mathrm{b} + n^\mathrm{Q}_\mathrm{b}.
\end{equation}

The isospin of baryons are denoted by $\bm{\tau}_i$. To describe the baryon-baryon interactions, the isoscalar-scalar meson $\sigma$, isoscalar-vector meson $\omega_{\mu}$, and isovector-vector meson $\bm{\rho}_{\mu}$ are introduced with
$m_{\sigma}(g_{\sigma i})$, $m_{\omega}(g_{\omega i})$, and $m_{\rho}(g_{\rho i})$ being their masses (coupling constants), respectively. For a system with time-reversal symmetry, the space-like components of the vector fields $\omega_{\mu}$ and
$\bm{\rho}_{\mu}$ vanish, leaving only the time components $\omega_0$ and $\bm{\rho}_0$. Meanwhile, charge conservation guarantees that only the $3$rd component $\rho_{0,3}$ in the isospin space survives. Note that for uniform dense matter $\sigma$, $\omega_0$ and $\rho_{0,3}$ are independent of the space coordinates, so that their space and time derivatives vanish.

In the quarkyonic phase, baryons and quarks coexist inside a same volume. Similar to the treatments of $\alpha$ clustering
inside nuclear matter in Refs.~\cite{Roepke2014_PRC90-034304, Xu2016_PRC93-011306}, we adopt a phenomenological baryon mass scaling to consider the effects of Pauli blocking and interactions between quarks and baryons, i.e.,
\begin{equation}
 m_i(n^\mathrm{Q}_\mathrm{b})=m_{0i}+B n^\mathrm{Q}_\mathrm{b}, \label{eq:Bmass}
\end{equation}
where $m_{0i}$~$(i=n,p)$ represents the baryon mass in vacuum and $B$ the interaction strength.

The quarks are treated as quasi-free particles with density dependent equivalent masses in the framework of equivparticle models~\cite{Peng1999_PRC61-015201, Peng2002_PLB548-189, Wen2005_PRC72-015204, Xia2014_PRD89-105027}, which is described by the Lagrangian density in Eq.~(\ref{eq:2.3}). Considering the interactions of linear confinement and leading-order perturbation, the quark mass scaling is determined by~\cite{Xia2014_PRD89-105027}
\begin{equation}\label{eq:2.12}
  m_{i}(n_\mathrm{b})=m_{0i}+\frac{D}{\sqrt[3]{n_\mathrm{b}}}+C\sqrt[3]{n_\mathrm{b}},
\end{equation}
where $m_{0u}=2.3$~MeV and $m_{0d}=4.8$~MeV are the current masses of quarks~\cite{Olive2014_CPC38-090001}. The parameter $D$ represents the confinement strength, which is related to the chiral restoration density, string tension, and the sum of vacuum chiral condensates. The perturbative strength $C$ is connected to the strong coupling constant. Due to the uncertainties in relevant quantities, the exact values of $D$ and $C$ are still unclear. Nevertheless, it has been estimated that $\sqrt{D}$ approximately lies in the range of $147$-$270$~MeV~\cite{Wen2005_PRC72-015204} and $C\lesssim 1.2$~\cite{Xia2014_PRD89-105027}.

According to the Typel-Wolter ansatz~\cite{Typel1999_NPA656-331}, we adopt density-dependent nucleon-meson coupling constants. For $\sigma$ and $\omega$ mesons, the coupling constants are determined by
\begin{equation}
\label{eq:2.5}
 g_{\phi i}(n^\mathrm{B}_\mathrm{b})=g_{\phi i}(n_{0}) a_{\phi}\frac{1+b_{\phi}(x+d_{\phi})^{2}}{1+c_{\phi}(x+e_{\phi})^{2}},
\end{equation}
where $\phi=\sigma, \omega$ and $x \equiv n^\mathrm{B}_\mathrm{b}/n_{0}$ with $n_{0}$ being the saturation density of nuclear matter. $a_{\phi}$, $b_{\phi}$, $c_{\phi}$, $d_{\phi}$, and $e_{\phi}$ are five adjustable parameters describing the density dependent coupling constants. Meanwhile, a different formula is adopted for $\rho$ meson, i.e.,
\begin{equation}
\label{eq:2.6}
 g_{\rho i}(n^\mathrm{B}_\mathrm{b})=g_{\rho i}(n_{0})  \mathrm{exp} [-a_{\rho}(x+b_{\rho})].
\end{equation}

Based on the Lagrangian density in Eqs.~(\ref{eq:2.2}-\ref{eq:2.4}), the meson fields are determined by
\begin{eqnarray}
\label{eq:2.7}
  m_{\sigma}^{2} \sigma&=&-\sum_{i=n,p} g_{\sigma i} n_{i}^{s},\\
\label{eq:2.8}
 m_{\omega}^{2} \omega_{0}&=&\sum_{i=n,p} g_{\omega i} n_{i},\\
\label{eq:2.9}
 m_{\rho}^{2} \rho_{3}&=&\sum_{i=n,p} g_{\rho i} \tau_{i,3} n_{i}.
\end{eqnarray}
Adopting no-sea approximation, the source currents of fermion $i$ for cold dense matter are given by
\begin{eqnarray}
\label{eq:2.10}
 n_{i}&=&\langle \bar{\Psi}_{i} \gamma^{0} \Psi_{i} \rangle = \frac{g_{i}\nu_{i}^{3}}{6\pi^{2}},\\
\label{eq:2.11}
n_{i}^{s}&=&\langle \bar{\Psi}_{i} \Psi_{i} \rangle = \frac{g_i (m_{i}^{*})^{3}}{4\pi^{2}}\left[x_{i}\sqrt{x_{i}^{2}+1}-\mathrm{arcsh}(x_{i})\right].
\end{eqnarray}
Here we have defined $x_{i} \equiv \nu_{i}/m_{i}^{*}$ with $\nu_{i}$ being the Fermi momentum and the degeneracy factor $g_{i}$ are taken as $g_{n,p}=2, g_{u,d}=6, g_{e,\mu}=2$ for baryons, quarks and leptons, respectively.
The effective mass for baryon $b$ is defined as $m_b^{*}=m_b(n^\mathrm{Q}_\mathrm{b})+g_{\sigma b} \sigma$ with the baryon mass scaling $m_b(n^\mathrm{Q}_\mathrm{b})$ indicated in Eq.~(\ref{eq:Bmass}), while for quark $q$ we adopt the mass scaling of Eq.~(\ref{eq:2.12}), i.e., $m_{q}^{*}=m_{q}(n_\mathrm{b})$. Meanwhile, the masses of leptons remain constant with $m_e^*=0.511~$MeV and $m_{\mu}^*=105.66~$MeV~\cite{Olive2014_CPC38-090001}. The single particle energies of fermions at fixed momentum $p$ are
\begin{eqnarray}
\epsilon_b^\mathrm{B}(p)&=&g_{\omega b} \omega+g_{\rho b} \tau_{b,3} \rho_{3}+\Sigma_b^\mathrm{R}+\sqrt{p^{2}+(m_b^{*})^{2}}, \label{eq:spl_b}\\
 \epsilon_{q}^\mathrm{Q}(p)&=&\Sigma_{q}^\mathrm{R}+\sqrt{p^{2}+(m_{q}^*)^{2}}, \label{eq:spl_q}\\
 \epsilon_{l}^\mathrm{L}(p)&=&\sqrt{p^{2}+(m_{l}^{*})^2}, \label{eq:spl_l}
\end{eqnarray}
with the ``rearrangement'' terms given by
\begin{eqnarray}
\label{eq:Sigma_b}
 \Sigma_b^\mathrm{R}&=&\sum_{i=n,p} \left(\frac{\mbox{d} g_{\sigma i}}{\mbox{d} n^\mathrm{B}_\mathrm{b}}\sigma n_{i}^s + \frac{\mbox{d} g_{\omega i}}{\mbox{d} n^\mathrm{B}_\mathrm{b}} \omega n_{i} + \frac{\mbox{d} g_{\rho i}}{\mbox{d} n^\mathrm{B}_\mathrm{b}} \rho_{3} \tau_{i,3} n_{i}\right) \nonumber\\
 &&+\sum_{i=u,d} \frac{\mbox{d} m_{i}}{\mbox{d} n_\mathrm{b}}n_{i}^s,\\
\label{eq:Sigma_Q}
 \Sigma_{q}^\mathrm{R}&=&\frac{1}{3}\sum_{i=n,p,u,d}\frac{\mbox{d} m_{i}}{\mbox{d} n_\mathrm{b}}n_{i}^s.
\end{eqnarray}

For quarkyonic matter, the quark-hadron interface in momentum space is set by matching the single particle energies, i.e.,
\begin{equation}
 \left\{\begin{array}{c}
   \epsilon_u^\mathrm{Q}(\nu_u) + 2\epsilon_d^\mathrm{Q}(\nu_d)=\epsilon_n^\mathrm{B}(0) \\
  2\epsilon_u^\mathrm{Q}(\nu_u) +  \epsilon_d^\mathrm{Q}(\nu_d)=\epsilon_p^\mathrm{B}(0) \\
 \end{array}\right. ,
\label{eq:Qy_spl}
\end{equation}
where $\epsilon_i(p_i)$ represents the single particle energy at a given momentum $p_i$. In the quarkyonic phase, $\nu_{u}$ and $\nu_{d}$ now represent the maximum momentums for $u$ and $d$ quarks instead of Fermi momentums, above which are baryons and the effects of Pauli blocking exclude the existence of free quarks. The chemical potentials for baryon $b$ and lepton $l$ are then fixed by $\mu_b = \epsilon_b^\mathrm{B}(\nu_b)$ and $\mu_l = \epsilon_l^\mathrm{L}(\nu_l)$. For quarks, we can also define an effective chemical potential $\mu_q = \epsilon_q^\mathrm{Q}(\nu_q)$, which is nonetheless not the actual one as $\nu_q$ does not correspond to the Fermi surface in the quarkyonic phase.

Finally, the energy density can be determined by
\begin{equation}
  E=\sum_{i} \varepsilon_{i} (\nu_{i},m_{i}^{*})+\sum_{\phi=\sigma, \omega, \rho} \frac{1}{2} m_{\phi}^{2} \phi^{2}, \label{eq:ener}
\end{equation}
with the kinetic energy density
\begin{eqnarray}
\label{eq:2.15}
 \varepsilon_{i} (\nu_{i},m_{i}^{*}) &=&\int_{0}^{\nu_{i}} \frac{g_{i} p^{2}}{2 \pi^{2}}\sqrt{p^{2}+(m_{i}^{*})^{2}}dp \\
     & =&\frac{g_{i}(m_{i}^{*})^{4}}{16 \pi^{2}}[x_{i}(2x_{i}^{2}+1)\sqrt{x_{i}^{2}+1}-\mathrm{arcsh}(x_{i})].\nonumber
\end{eqnarray}
Then the pressure $P$ is obtained with
\begin{equation}
 P=\sum_{i} \mu_{i} n_{i}-E. \label{eq:pressure}
\end{equation}

At a given total baryon number density $n_\mathrm{b}$ and isospin asymmetry
\begin{equation}
  \delta\equiv (n_d-n_u+n_n-n_p)/n_\mathrm{b},
\end{equation}
the properties of three types of strongly interacting matter can then be fixed, i.e.,
\begin{enumerate}
  \item Nuclear matter: $n^\mathrm{Q}_\mathrm{b}=0$ with $n_\mathrm{b}= n^\mathrm{B}_\mathrm{b}$;
  \item Quark matter: $n^\mathrm{B}_\mathrm{b}=0$ with $n_\mathrm{b}= n^\mathrm{Q}_\mathrm{b}$;
  \item Quarkyonic matter: $n_\mathrm{b}= n^\mathrm{B}_\mathrm{b}+n^\mathrm{Q}_\mathrm{b}$ with $n^\mathrm{B}_\mathrm{b}$ and $n^\mathrm{Q}_\mathrm{b}$ fixed by Eq.~(\ref{eq:Qy_spl}).
\end{enumerate}
The corresponding mean fields, single particle energies, and densities are fixed by solving Eqs.~(\ref{eq:Bmass}-\ref{eq:Qy_spl}) in an iterative manner. Once convergency is reached, the energy density and pressure can then be obtained with Eqs.~(\ref{eq:ener}) and (\ref{eq:pressure}).

\section{\label{sec:res}Results and discussions}

\begin{table}
\caption{\label{table:NM} Saturation properties of nuclear matter predicted by three different density-dependent covariant density functionals TW99~\cite{Typel1999_NPA656-331}, PKDD~\cite{Long2004_PRC69-034319}, and DD-ME2~\cite{Lalazissis2005_PRC71-024312}. }
\begin{tabular}{c|ccccccc} \hline \hline
       & $n_0$        &   $B$    &   $K$  &  $J$   & $S$    &  $L$  & $K_\mathrm{sym}$        \\
       & fm${}^{-3}$  &   MeV    &   MeV  &  MeV   &  MeV   &  MeV  &   MeV             \\ \hline
TW99   &  0.153       & $-$16.24 &  240.2 & $-540$ & 32.8   &  55.3 &  $-125$          \\
PKDD   &  0.150       & $-$16.27 &  262.2 & $-119$ & 36.8   &  90.2 &  $-81$           \\
DD-ME2 &  0.152       & $-$16.13 &  250.8 & 477    & 32.3   &  51.2 &  $-87$           \\
\hline
\end{tabular}
\end{table}

For baryonic matter described by the Lagrangian density in Eq.~(\ref{eq:2.2}), we adopt three different density-dependent covariant density functionals TW99~\cite{Typel1999_NPA656-331}, PKDD~\cite{Long2004_PRC69-034319}, and DD-ME2~\cite{Lalazissis2005_PRC71-024312}. The corresponding properties of nuclear matter around the saturation density ($n_0\approx 0.16\ \mathrm{fm}^{-3}$) are indicated in Table~\ref{table:NM}, which include the binding energy $B$, incompressibility $K$, skewness coefficient $J$, symmetry energy $S$, slope $L$ and curvature parameter $K_\mathrm{sym}$ of nuclear symmetry energy. Note that some of the coefficients are well constrained with $B\approx -16$ MeV, $K = 240 \pm 20$ MeV~\cite{Shlomo2006_EPJA30-23}, $S = 31.7 \pm 3.2$ MeV, and $L = 58.7 \pm 28.1$ MeV~\cite{Li2013_PLB727-276, Oertel2017_RMP89-015007}, which can be further constrained by considering the recent data from astrophysical observations, heavy-ion collisions, measurements of the neutron skin thicknesses, and nuclear theories~\cite{Zhang2020_PRC101-034303, Xie2021_JPG48-025110, PREX2021_PRL126-172502, Essick2021_PRL127-192701, CREX2022_PRL129-042501}. The saturation properties of nuclear matter predicted by the covariant density functionals generally coincide the those constraints, except that PKDD predicts slightly larger $S$ and $L$. In summary, compared with TW99, the functional PKDD predicts larger symmetry energy ($S$ and $L$), while the energy per baryon for symmetric nuclear matter at supra-saturation densities is significantly increased (larger $K$ and $J$) if DD-ME2 is adopted.

\begin{table}
  \centering
  \caption{\label{table:QyM} The adopted parameter sets $(B, C, \sqrt{D})$ for the baryon and quark mass scalings in Eqs.~(\ref{eq:Bmass}) and (\ref{eq:2.12}). The obtained radii $R_{1.4}$ and tidal deformability $\Lambda_{1.4}$ of 1.4-solar-mass compact stars, the maximum mass $M_\mathrm{TOV}$, and the maximum sound speed $v_\mathrm{max}$ of quarkyonic matter are indicated as well.}
  \begin{tabular}{c|ccc|cccc}
    \hline \hline
         &$B$           & $C$ & $\sqrt{D}$ &$R_{1.4}$ &$\Lambda_{1.4}$ &$M_\mathrm{TOV}$ & $v_\mathrm{max}$ \\
         & MeV/fm$^{3}$ &     & MeV        & km       &                & $M_\odot$       & $c$         \\ \hline
\multirow{4}{*}{TW99}
         & {300}& {0.7} & {180}
                          &12.27 & 405 & 2.04 & 0.73\\ 
         & 0  & 0.7 & 180 &12.27 & 405 & 1.97 & 0.67\\
         & 300& 0.2 & 180 &12.20 & 386 & 1.88 & 0.68\\
         & 300& 0.7 & 230 &12.27 & 405 & 2.08 & 0.83\\ \hline
\multirow{4}{*}{PKDD}
         &\textbf{150} & \textbf{0.7} & \textbf{150} &12.80 & 530 & 2.06 &0.67 \\ \cline{2-4}
         & 0  & 0.7 & 150 &12.40 & 463 & 2.00 &0.67 \\
         & 150& 1.0 & 150 &13.60 & 751 & 2.20 &0.70 \\
         & 150& 0.7 & 180 &13.63 & 764 & 2.20 &0.69 \\    \hline
\multirow{4}{*}{DD-ME2}
         & \textbf{100}& \textbf{0.5} & \textbf{160} &12.74 & 557 & 2.06 &0.65 \\  \cline{2-4}
         & 300& 0.5 & 160 &13.08 & 666 & 2.15 &0.65 \\
         & 100& 0.7 & 160 &13.17 & 703 & 2.19 &0.65 \\
         & 100& 0.5 & 180 &13.20 & 712 & 2.19 &0.63 \\
    \hline \hline
  \end{tabular}
\end{table}

Based on the aforementioned density functionals, we further consider the possible formation of quarkyonic matter by including explicitly quasi-free quarks. The adopted parameter sets $(B, C, \sqrt{D})$ of the baryon and quark mass scalings in Eqs.~(\ref{eq:Bmass}) and (\ref{eq:2.12}) are listed in Table~\ref{table:QyM}, where $B$ is in MeV$/$fm$^{3}$, $C$ dimensionless, and $\sqrt{D}$ in MeV. To fix the properties of dense stellar matter, leptons fulfilling charge neutrality condition need to be considered, i.e.,
\begin{equation}\label{eq:2.22}
  \sum_{i} q_{i} n_{i}=0,
\end{equation}
where $q_{n}=0$, $q_{p}=1$, $q_{u}=2/3$, $q_{d}=-1/3$, and $q_{e}=q_{\mu}=-1$ are the charge number of each particle type. Note that hyperons are not included yet, which will be considered in our future works. Additionally, at fixed total baryon number density $n_\mathrm{b}$, the number densities of leptons $n_{e, \mu}$, quarks $n^\mathrm{Q}_\mathrm{b}$, and isospin asymmetry $\delta$ for cold dense stellar matter are fixed by fulfilling the chemical equilibrium condition, i.e.,
\begin{equation}\label{eq:2.23}
  \mu_{n} - \mu_{p} = \mu_{e}=\mu_{\mu}.
\end{equation}
The EOSs of neutron star matter are obtained with the energy density $E$ fixed by Eq.~(\ref{eq:ener}) and pressure $P$ by Eq.~(\ref{eq:pressure}).

\begin{figure}[!ht]
  \centering
  \includegraphics[width=0.95\linewidth]{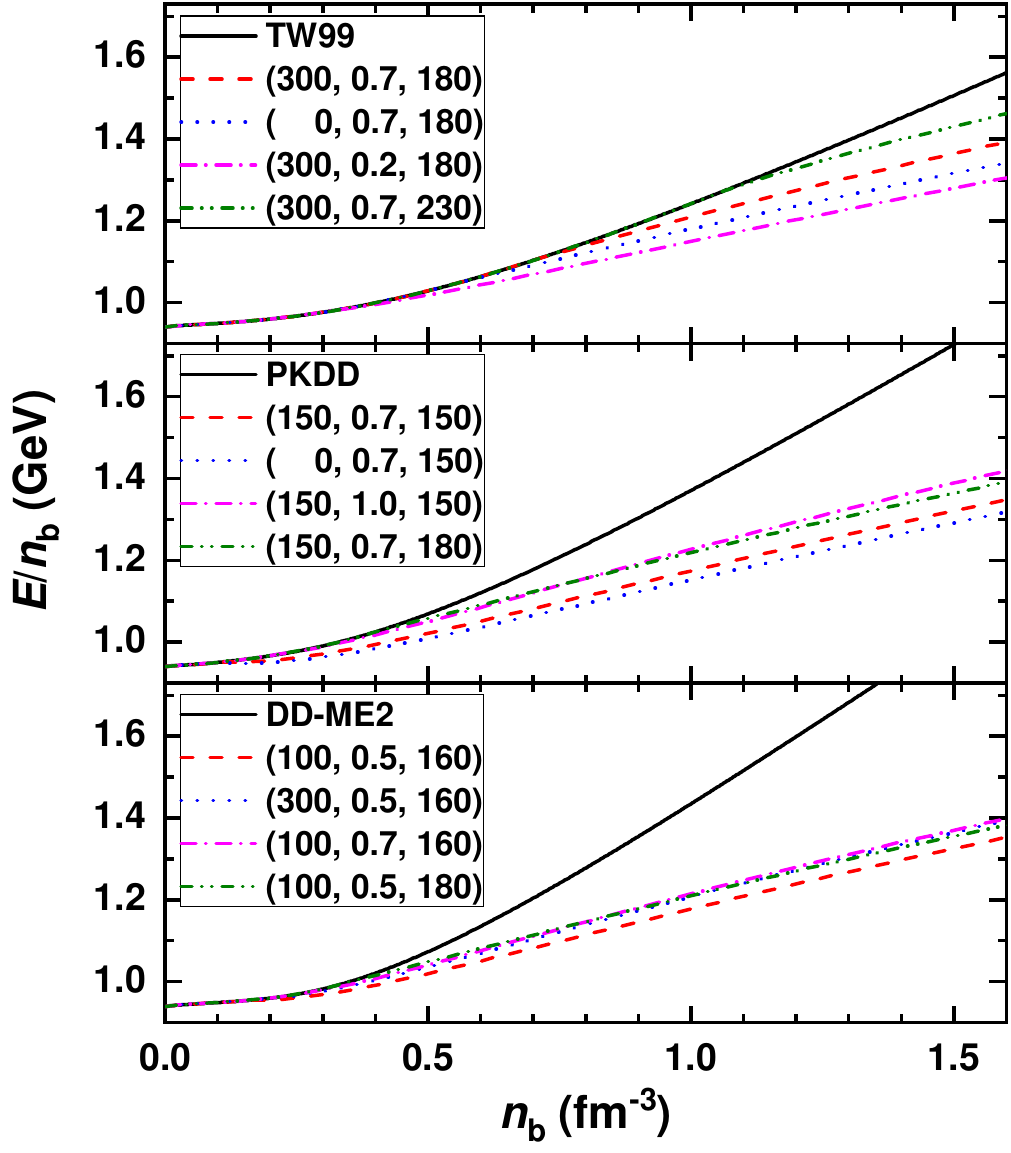}
  \caption{\label{Fig:EpAall}Energy per baryon $E/n_\mathrm{b}$ of nuclear matter (solid lines) and quarkyonic matter (dashed lines) as functions of the total baryon number density $n_\mathrm{b}$, which are obtained adopting the parameter sets indicated in Table.~\ref{table:QyM}.}
\end{figure}

In Fig.~\ref{Fig:EpAall}, we present the energy per baryon $E/n_\mathrm{b}$ of nuclear matter and quarkyonic matter in compact stars as functions of the total baryon number density $n_\mathrm{b}$. As density increases, nuclear matter are converted into quarkyonic matter at $n_\mathrm{b}\gtrsim 0.1\ \mathrm{fm}^{-3}$, which becomes more stable as the energy is decreased. The variations of the energy per baryon in quarkyonic matter are far less significant than that of nuclear matter. It is found that the transitions from nuclear matter to quarkyonic matter are mostly second-order, except for the case of PKDD adopting the parameter set  $B=0$, $C=0.7$, and $\sqrt{D}=150$ MeV, where a first-order quarkyonic transition is identified. The effects of various types of interactions can be examined by varying the corresponding parameters, where the onset densities of quarkyonic transitions and energies of quarkyonic matter increase with the strengths of quark-hadron interaction $B$, perturbative interaction $C$, and confinement $\sqrt{D}$. Meanwhile, we note that increasing $C$ leads to more significant increment in energy at higher densities, which is mainly due to the increasing repulsive interaction described by the quark mass scaling in Eq.~(\ref{eq:2.12}).

\begin{figure}[!ht]
\centering
  \includegraphics[width=0.95\linewidth]{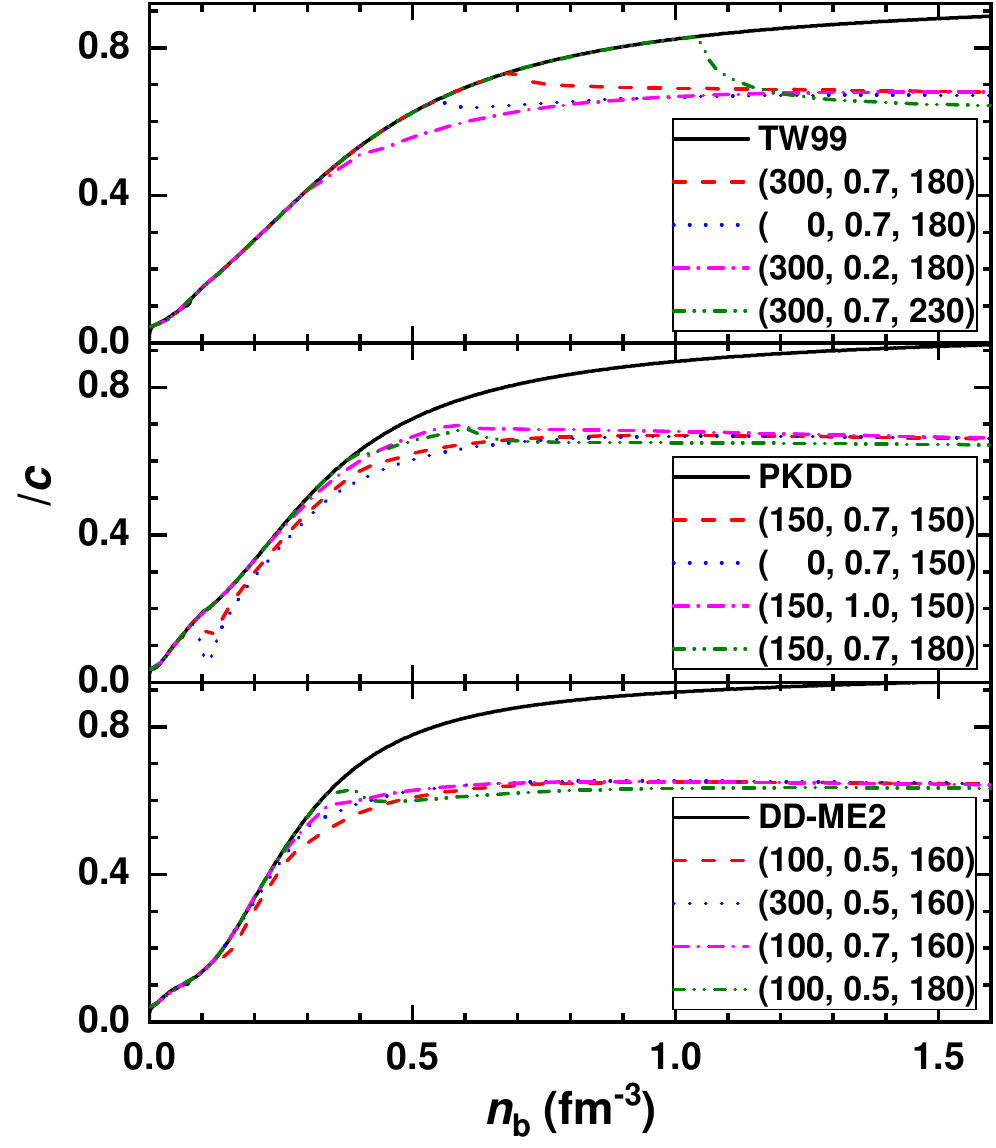}
  \caption{\label{Fig:ALLV} Velocity of sound $v$ in nuclear matter (solid lines) and quarkyonic matter (dashed lines) obtained with the EOSs presented in Fig.~\ref{Fig:EpAall}.}
\end{figure}

To show more explicitly the variations in the stiffness of the EOSs, we present the velocity of sound $v$ in Fig.~\ref{Fig:ALLV}, which is determined by
\begin{equation}
\label{eq:v}
v=\sqrt{\frac{dP}{dE}}.
\end{equation}
As the total baryon number density $n_\mathrm{b}$ increases, the velocity of sound also increases before reaching its peak $v_\mathrm{max}$ for quarkyonic matter. Such structure in the speed of sound was identified in various previous studies and interpreted as the onset of a new matter state~\cite{Annala2020_NP, Ma2020_PPNP113-103791, Jin2022_PLB829-137121, Tan2022_PRL128-161101, Han2023_SB68-913}, which corresponds to quarkyonic transition in our current study. At larger densities, the velocity of sound for quarkyonic matter is distinctively smaller than that of nuclear matter, which approaches to $\sim$0.64 and is slightly larger than the ultrarelativistic limit 1/$\sqrt{3}$ ($\approx 0.58$). Note that when we take $B= 0$, $C= 0.7$ and $\sqrt{D}= 150$~MeV for PKDD, the quarkyonic transition is of first-order and the velocity of sound is zero in the range $n_\mathrm{b}\approx 0.1$ to 0.13 fm$^{-3}$. Meanwhile, at large enough densities, e.g., $\sim$$40n_0$, perturbative QCD is applicable and we expect the formation of a deconfined quark matter with $v\rightarrow 1/\sqrt{3}$~\cite{Freedman1977_PRD16-1169, Fraga2005_PRD71-105014, Kurkela2010_PRD81-105021}. The deviation of $v$ from $1/\sqrt{3}$ is thus attributed to the strong interactions in the quaryonic phase. Generally speaking, at small densities with the emergence of quarkyonic matter, the velocity of sound increases with $B$, $C$, and $\sqrt{D}$, which can be identified as well according to the maximum sound speed $v_\mathrm{max}$ indicated in Table~\ref{table:QyM}. At larger densities, the velocity of sound increases with $C$ and decreases with $\sqrt{D}$, while varying the quark-hadron interaction strength $B$ has little contribution to $v$.

Based on the EOSs presented in Fig.~\ref{Fig:EpAall}, the corresponding structures of compact stars are obtained by solving the Tolman-Oppenheimer-Volkov (TOV) equation
\begin{equation}
\frac{\mbox{d}P}{\mbox{d}r} = -\frac{G M E}{r^2}   \frac{(1+P/E)(1+4\pi r^3 P/M)} {1-2G M/r}  \label{eq:2.28}
\end{equation}
with the subsidiary condition
\begin{equation}
\frac{\mbox{d}M}{\mbox{d}r} = 4\pi E r^2 \label{eq:2.29}.
\end{equation}
The gravity constant is taken as $G=6.707\times 10^{-45}\ \mathrm{MeV}^{-2}$. The dimensionless tidal deformability is calculated by
\begin{equation}
\Lambda = \frac{2 k_2}{3}\left( \frac{R}{G M} \right)^5, \label{eq:2.30}
\end{equation}
where the second Love number $k_2$ is evaluated by introducing perturbations to the metric~\cite{Damour2009_PRD80-084035, Hinderer2010_PRD81-123016, Postnikov2010_PRD82-024016}. Note that a first-order liquid-gas phase transition takes place at subsaturation densities, which forms various types of nonuniform structures and we have adopted unified neutron star EOSs corresponding to the employed covariant density functionals~\cite{Xia2022_CTP74-095303}.

\begin{figure*}[!ht]
  \centering
  \includegraphics[width=0.75\linewidth]{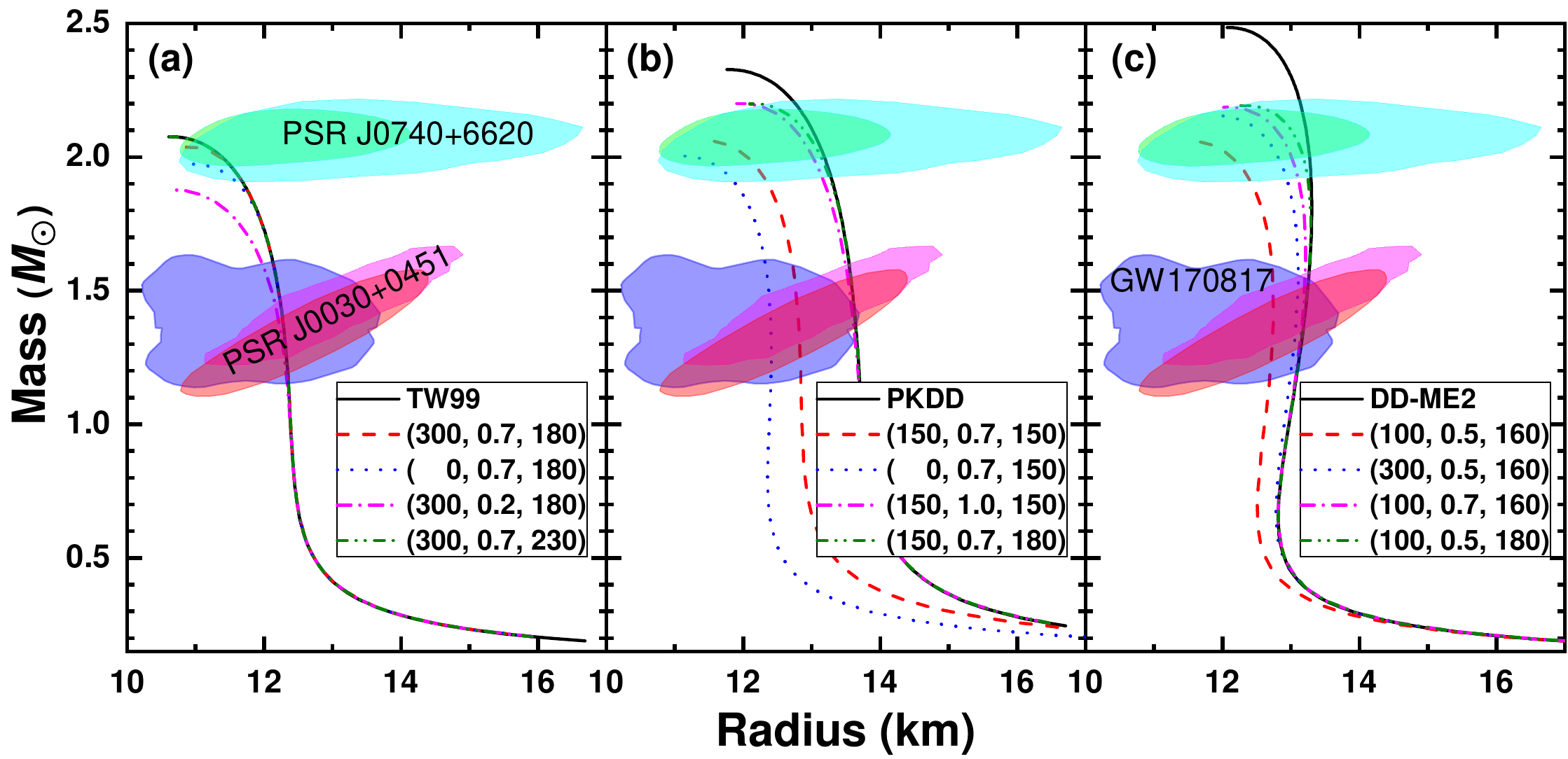}
\caption{\label{Fig:MR} Mass-radius relations of compact stars obtained with the EOSs presented in Fig.~\ref{Fig:EpAall}. The shaded regions indicate the constraints from the binary neutron star merger event GW170817 within 90\% credible region~\cite{LVC2018_PRL121-161101}, the observational pulse-profiles in PSR J0030+0451 and PSR J0740+6620 within 68\% credible region~\cite{Riley2019_ApJ887-L21, Riley2021_ApJ918-L27, Miller2019_ApJ887-L24, Miller2021_ApJ918-L28}.}
 \label{Fig:MRall}
\end{figure*}

In Fig.~\ref{Fig:MR} we present the $M$-$R$ relations of neutron stars and quarkyonic stars obtained by adopting different combinations of parameters in Table~\ref{table:QyM}. The corresponding radius $R_{1.4}$ and tidal deformability $\Lambda_{1.4}$ for $1.4M_\odot$ stars, and the maximum mass $M_{\rm TOV}$ are indicated in Table~\ref{table:QyM} as well. Based on various observational data of pulsars, strong constraints on compact star structures are obtained. For example, by analyzing the orbital motion of pulsars in a binary system~\cite{Lattimer2012_ARNPS62-485}, the masses of PSR J1614-2230 (1.928 $\pm$ 0.017 $M_{\odot}$)~\cite{Fonseca2016_ApJ832-167} and PSR J0348+0432 (2.01 $\pm$ 0.04 $M_{\odot}$)~\cite{Antoniadis2013_Science340-1233232} were measured with high precision. The observation of gravitational waves emitted in the binary neutron star merger event GW170817 has placed strong constraints on the tidal deformability 70 $\leq \Lambda_{1.4} \leq$ 580, corresponding to a radius of 11.9$_{-1.4}^{+1.4}$~km~\cite{LVC2018_PRL121-161101}. The simultaneous measurements of masses and radii for PSR J0030+0451 and PSR J0740+6620 have also placed strong constraints on compact star structures~\cite{Riley2019_ApJ887-L21, Riley2021_ApJ918-L27, Miller2019_ApJ887-L24, Miller2021_ApJ918-L28}.

The $M$-$R$ relation of neutron stars predicted by the covariant density functional TW99 agrees well with the observational constraints~\cite{Xia2022_CTP74-095303}, while the radii for two-solar-mass neutron stars lie in the lower ends of the PSR J0740+6620 constraints~\cite{Riley2021_ApJ918-L27, Miller2021_ApJ918-L28}. Nevertheless, neutron stars obtained with PKDD (larger $L$) and DD-ME2 (larger $J$) have larger maximum masses $M_{\rm TOV}$,  radii $R_{1.4}$, and tidal deformabilities $\Lambda_{1.4}$, where $R_{1.4}$ and $\Lambda_{1.4}$ slightly exceed the observational upper limits. With the emergence of quarkyonic matter, the EOSs of quaryonic matter become softer and consequently quarkyonic stars are more compact with smaller radii and tidal deformabilities. For smaller values of ($B$, $C$, $\sqrt{D}$), the EOSs of quaryonic matter become softer, where $M_{\rm TOV}$, $R_{1.4}$, and $\Lambda_{1.4}$ decrease. The quarkyonic stars obtained with the parameter sets (100, 0.5, 160) for DD-ME2, (150, 0.7, 150) and (0, 0.7, 150) for PKDD thus become consistent with various constraints from pulsar observations. Note that for TW99, quarkyonic transition is not favored according to pulsar observations, where quarkyonic matter can only emerge in the center regions of massive stars. Evidently, adopting (300, 0.2, 180) for TW99 predicts a too soft EOS for quaryonic stars, where the corresponding maximum mass does not reach 2$M_{\odot}$ and is thus inconsistent with pulsar observations~\cite{Antoniadis2013_Science340-1233232}. In such cases, quarkyonic transition is more likely to take place if a large skewness coefficient $J$ or slope of symmetry energy $L$ is confirmed for nuclear matter, e.g., those from PREX-2~\cite{PREX2021_PRL126-172502}.

\begin{figure*}[htbp]
\begin{minipage}[t]{0.5\linewidth}
\centering
\includegraphics[width=\linewidth]{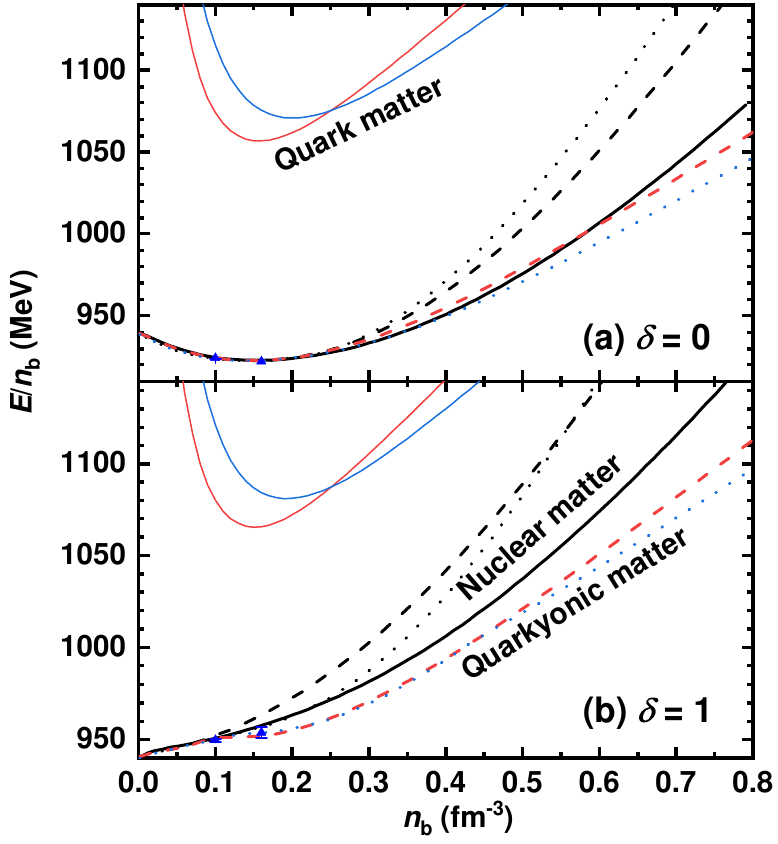}
\end{minipage}%
\hfill
\begin{minipage}[t]{0.475\linewidth}
\centering
\includegraphics[width=\linewidth]{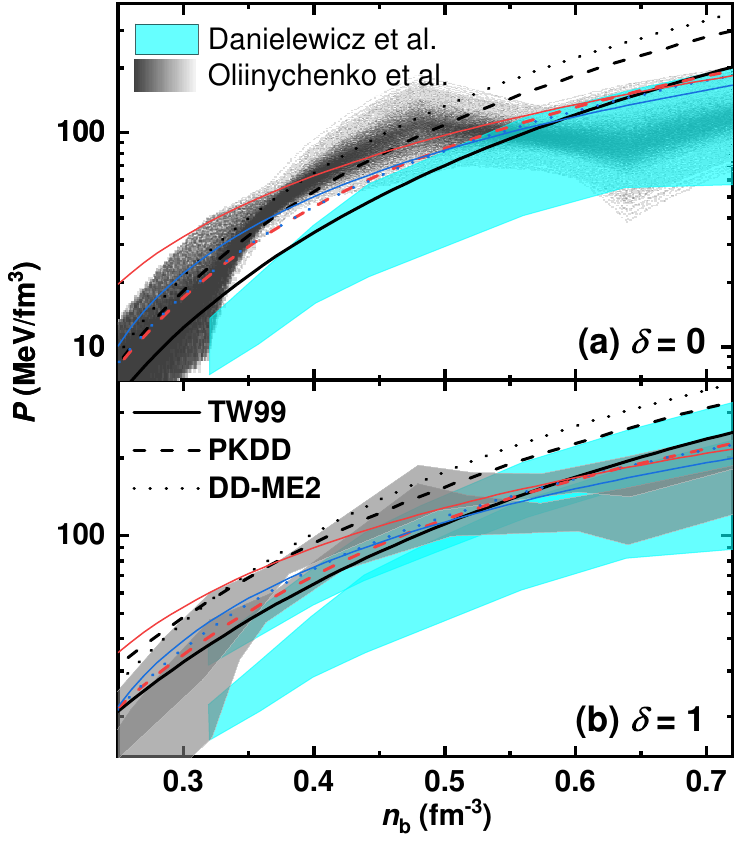}
\end{minipage}
\caption{\label{Fig:EOS_fix} Energy per baryon and pressure of nuclear matter (black), quark matter (color-solid), and  quarkyonic matter (color-dashed) with isospin asymmetry~(a)~$\delta=0$ and (b)~$\delta=1$. The solid triangles in the left panel indicate the constrains from finite nuclei properties~\cite{Brown2013_PRL111-232502, Li2013_PLB727-276, Oertel2017_RMP89-015007}, while the color bands in the right panel correspond to the constraints from the experimental flow data~\cite{Danielewicz2002_Science298-1592, Oliinychenko2022}.}
\end{figure*}

In Fig.~\ref{Fig:EOS_fix}, we present energy per baryon and pressure of nuclear matter, quark matter, and quarkyonic matter with isospin asymmetry $\delta=0$ and 1. To fix the properties of quark matter and quarkyonic matter, as indicated by the boldface in Table.~\ref{table:QyM}, the parameter sets (100, 0.5, 160) for DD-ME2 and (150, 0.7, 150) for PKDD are adopted, which predict quarkyonic stars that are consistent with pulsar observations. Evidently, the deconfined quark matter is highly unstable in comparison with nuclear matter and quarkyonic matter, where the energy per baryon is much larger. The quarkyonic transition takes place at around 1.4$n_0$ for symmetric nuclear matter~(SNM), while the onset density is decreased significantly for pure neutron matter~(PNM) at around $n_\mathrm{on} = 0.1 \ \mathrm{fm}^{-3}$. Similar to quarkyonic matter in compact stars, the energy is decreased once quarkyonic transition takes place for both SNM and PNM. The obtained energy per baryon is then compared with the well-constrained nuclear matter properties at $n_0$ and $n_\mathrm{on}$, i.e., $B(n_0) = -16$ MeV, $S(n_0) = 31.7 \pm 3.2$ MeV~\cite{Li2013_PLB727-276, Oertel2017_RMP89-015007}, $B(n_\mathrm{on}) = -14.1\pm0.1$ MeV, and $S(n_\mathrm{on})=25.5\pm1.0$ MeV~\cite{Centelles2009_PRL102-122502, Brown2013_PRL111-232502}. Evidently, the binding energy of SNM agrees well with the constraints $B(n_\mathrm{on}) = -14.1\pm0.1$ MeV and $B(n_0) = -16$ MeV. This is not the case for PNM, where PKDD predicts symmetry energy that exceeds the constraint $S(n_0) = 31.7 \pm 3.2$ MeV. The situation is improved if quarkyonic transition takes place for PNM, which well reproduce the constraint on symmetry energy $S(n_0) = 31.7 \pm 3.2$ MeV.

In the right panel of Fig.~\ref{Fig:EOS_fix}, we compare the pressure of nuclear matter, quark matter, and quarkyonic matter  with various constraints from the flow data of heavy-ion collisions~\cite{Danielewicz2002_Science298-1592, Oliinychenko2022}. Note that there exist many other constraints on the pressure of dense matter~\cite{Fuchs2004_PPNP53-113, Lynch2009_PPNP62-427, LeFevre2016_NPA945-112}, which are not indicated in Fig.~\ref{Fig:EOS_fix} since they generally coincide with those from Ref.~\cite{Danielewicz2002_Science298-1592}. For SNM at $n_\mathrm{b}\approx 2$-3$n_0$, the pressure obtained by RMF models are generally larger than the constraint provided by Danielewicz et al.~\cite{Danielewicz2002_Science298-1592}, which nonetheless coincide with the constraint from Oliinychenko et al.~\cite{Oliinychenko2022}. At larger densities, however, SNM becomes too stiff except for those obtained with the covariant energy density functional TW99. This can be improved if we consider quarkyonic transitions, where the pressure at $n_\mathrm{b}\approx 3$-5$n_0$ coincide with the constraints from the flow data of heavy-ion collisions~\cite{Danielewicz2002_Science298-1592, Oliinychenko2022}. Similar situations are also observed for PNM, where the pressure obtained with PKDD and DD-ME2 are too large except for TW99 that gives satisfactory results. Note that at $n_\mathrm{b}\approx 2$-3$n_0$, the constraint from Oliinychenko et al.~\cite{Oliinychenko2022} gives larger upper limit on pressure as well, which supports the predictions of RMF models. At larger densities, it is necessary to consider quarkyonic transitions if the covariant density functionals PKDD and DD-ME2 are adopted, where the pressure of PNM is decreased so that it is consistent with the constraints from heavy-ion collisions~\cite{Danielewicz2002_Science298-1592}.

\section{\label{sec:con}Conclusion}

In this work, by combining RMF models and equivparticle models with density-dependent quark masses~\cite{Xia2018_JPSCP20-011010}, we extend RMF models to include quark degrees of freedom, where we have constructed explicitly ``a quark Fermi Sea'' and ``a baryonic Fermi surface'' to model the quarkyonic phase. In contrast to previous treatments of simply removing lower momentum components~\cite{McLerran2019_PRL122-122701, Margueron2021_PRC104-055803, Cao2020_JHEP10-168, Cao2022_PRD105-114020}, baryons with momentums ranging from zero to Fermi momentums are included in our approach, which are more reasonable in  analogous to the formation of Cooper pairs that are dominated by zero momentum components. The nuclear matter, quark matter, and quarkyonic matter are treated in a unified manner. As we increase the density of nuclear matter, quarkyonic matter emerge and the energy per baryon decreases, i.e., quarkyonic matter is more stable than nuclear matter or quark matter and energy minimization is still applicable to obtain the microscopic properties of quarkyonic matter.

We have adopted three different effective baryon-baryon interactions TW99~\cite{Typel1999_NPA656-331}, PKDD~\cite{Long2004_PRC69-034319}, and DD-ME2~\cite{Lalazissis2005_PRC71-024312}, which indicates different saturation properties for nuclear matter with larger slope of symmetry energy $L$ for PKDD and larger skewness coefficient $J$ for DD-ME2 in comparison with TW99. Note that the covariant density functional TW99 gives satisfactory predictions for the nuclear matter properties both in neutron stars and heavy-ion collisions, where the quarkyonic transition is unfavorable according to both experimental and astrophysical constraints. This is not the case for either PKDD or DD-ME2, which predicts too stiff EOSs for nuclear matter in neutron stars and heavy-ion collisions. The radii and tidal deformabilities of neutron stars are too large with $R_{1.4} = 13.63$ km and $\Lambda_{1.4}=764$ for PKDD and $R_{1.4} = 13.2$ km and $\Lambda_{1.4}=712$ for DD-ME2, which exceeds the constraints $70\leq \Lambda_{1.4}\leq 580$ from the binary neutron star merger event GW170817~\cite{LVC2019_PRL123-161102} and the radius measurements of PSR J0030+0451 with $R_{1.4}=12.45\pm 0.65$ km~\cite{Miller2021_ApJ918-L28}. Meanwhile, the functionals PKDD and DD-ME2 predict too large pressure for nuclear matter at $n_\mathrm{b}\approx 3$-5$n_0$ according to the constraints from the flow data of heavy-ion collisions~\cite{Danielewicz2002_Science298-1592, Oliinychenko2022}. This situation can be improved if quarkyonic transition takes place, where the EOSs become softer and can accommodate various experimental and astrophysical constraints.

\begin{acknowledgments}
C.-J.X. would like to thank Prof. Yi-Zhong Fan, Prof. Sophia Han, and Dr. Yong-Jia Huang for fruitful discussions.
This work was partly supported by the National Natural Science Foundation of
China (Grant Nos. U2032141 and 12275234), the National SKA Program of
China (No. 2020SKA0120300), the Natural Science Foundation of Henan Province
(202300410479), the Foundation of Fundamental Research for Young Teachers of
Zhengzhou University (JC202041041), and the Physics Research and Development
Program of Zhengzhou University (32410217).
\end{acknowledgments}



%

\end{document}